\documentclass{JHEP3}

\def\makeatletter{\catcode`\@=11}% 11:letter
\makeatletter
\def\mathbox#1{\hbox{$\m@th#1$}}%
\def\math@ccstyles#1#2#3#4#5#6#7{{\leavevmode
      \setbox0\mathbox{#6#7}%
      \setbox2\mathbox{#4#5}%
      \dimen@ #3%
      \baselineskip\z@\lineskiplimit#1\lineskip\z@
      \vbox{\ialign{##\crcr
             \hfil \kern #2\box2 \hfil\crcr
             \noalign{\kern\dimen@}%
             \hfil\box0\hfil\crcr}}}}
\def\mathaccstyles{\math@ccstyles\maxdimen}
\def\maththroughstyles{\math@ccstyles{-\maxdimen}}
\def\unity%
 {\maththroughstyles{.45\ht0}\z@\displaystyle {\mathchar"006C}\displaystyle 1}

\title{M-Theory Brane as Giant Graviton and the Fractional Quantum Hall Effect}
\author{Ran Huo\\
        Shanghai Institute for Advanced Studies, USTC\\
        99 Xiupu Road, Nanhui, Shanghai, 201315, P.R. China\\
        and\\
        Department of Modern Physics, USTC\\
        Hefei, Anhui, 230026, P.R. China\\
        E-mail:\enspace\email{huor@ustc.edu}}

\abstract{A small number of M-theory branes as giant gravitons in
the M-theory sector of LLM geometry is studied as a probe. The
abelian way shows that the low energy effective action for M-theory
brane is exactly the 2d electron subject to a vertical magnetic
field. We also briefly discuss the microscopic description of
M2-brane giant graviton in this geometry, in the language of a
combination of D0-branes as fuzzy 2-spheres. Then we go to the
well-established Noncommutative Chern-Simons theory description.
After quantization, well behaved Fractional Quantum Hall Effect is
demonstrated. This goes beyond the original LLM description and
should be some indication of novel geometry.}


\preprint{SIAS-CMTP-06-4}

\begin{document}

\section{Introduction}
Counting fundamental degrees of freedom of theoretical physics
system remains a long term problem for theorists. In the brilliant
work started by J. Maldacena which is known as AdS/CFT
\cite{Maldacena}, a connection between the Supersymmetric Yang-Mills
theory which lives on branes and the supergravity (which can be
promoted to type IIB string/M-theory) on the asymptotic $AdS_m\times
S^n$ is established. Recently, the extension \cite{LLM} (known as
LLM) brings the whole $\frac{1}{2}$ BPS sector to our scope, in
which the original AdS/CFT correspondence can be viewed as a special
case. Motivated by the gauge theory side work \cite{Berenstein}, it
is shown that all the nonsingular $\frac{1}{2}$ BPS configurations
of the supergravity side that preserve certain symmetry can be
described by a black-white 2d moduli space, which can be viewed as
the 2d electron gas distribution of the Quantum Hall Effect. Once
the distribution is given, the geometry which preserves half the
supersymmetry is then uniquely determined by a Laplace/Toda
equation.\\
\indent As is known to all, the most mysterious aspect of the 2d
electron gas when exposed to strong vertical magnetic field is the
Fractional Quantum Hall Effect (FQHE) \cite{QHE}, which can be
naively regarded as the composite particle carries only fraction
charge while averaged to each parton. In \cite{refer}, the origin of
FQHE states in the LLM geometry is explored in the IIB sector, where
a small number of giant gravitons are treated as probe. As mentioned
in \cite{LLL}, the FQHE states in the AdS/CFT/QH formulism exceed
the $\frac{1}{2}$ BPS description, which deserves further study.\\
\indent Giant graviton \cite{GG1,GG2} plays an essential role in our
treatment. The original ansatz adopted by LLM preserve the symmetry
of $SO(4)\times SO(4)$ in the IIB sector and $SO(3)\times SO(6)$ in
the M-Theory sector, which are exactly the symmetries preserved by
both the internal and outer space giant gravitons in their separated
sectors. This reflects the consideration that it is the condensation
and interaction of giant gravitons that gives the novel LLM
geometry. Giant graviton functions as element for constructing
geometry, so it makes sense to separate a small number of giant
gravitons as a probe and take others as background. This separation
shed light on the possibility that is not covered by original LLM,
especially, the Fractional QHE.\\
\indent In the present paper, we will generalize \cite{refer} to the
M-theory brane case. According to \cite{M-born, BFSS}, M-theory is
the strong coupling limit of type IIA superstring theory, which
takes the 11d supergravity as its low energy effective theory.
M-theory has two kinds of branes, the M2-brane and the M5-brane,
they couple electrically and magnetically to the 3-form gauge
potential in the 11d supergravity. An exact microscopic description
of M-theory is still unknown, but for the M2-brane case, there is a
promising suggestion known as matrix model which suggests that the
M2-brane is obtained by a combination of large (infinite) number of
D0-branes \cite{BFSS,DLCQ1}. Although the original BFSS Matrix Model
\cite{BFSS} is formulated in the flat spacetime, some further work
show that the certain weakly coupled background case can also be
covered \cite{DLCQ1,curvedM1,curvedM2,curvedM3}. In \cite{curvedM3}
a set of D0-branes are combined into a fuzzy 2-sphere as a the
M2-brane giant graviton in the $AdS_7\times S^4$ and $AdS_4\times
S^7$ geometry, giving a microscopic description of M2-brane, which
is also a nontrivial extension of BFSS Matrix Model Conjecture. This
work shed light on our case, we will use similar technique in our
treatment.\\
\indent The paper is organized as follows. We present a brief review
of giant graviton in AdS/CFT and the LLM geometry in section 2,
emphasizing the M-theory sector. In section 3 we will show that the
low energy effective action of several coincident M-theory branes
are indeed governed by Quantum Hall Effect. First we use the
macroscopic (or abelian) way of directly computing the induced
metric and the pullback of the gauge potential for a single brane.
Then some brief comments will also be made about the microscopic (or
nonabelian) side, i.e., towards the DLCQ matrix model in a weakly
coupled background, in which we extend it to multi-brane case. In
the 4th section we will make use of the alternative (but somewhat
standard) description of QHE, i.e. the noncommutative Chern-Simons
theory (NCCS). We focus on the finiteness of giant gravitons and
revise the theory for consistency. After quantization we will
finally show the inverse filling factor quantization, i.e., the
emergence of the FQHE states. Then a brief discussion of the
interaction between the giant gravitons is presented. In the final
section we make some explanation and discussion of the meaning of
the solution and then conclude.

\section{Review of Giant Graviton and LLM Geometry}

\subsection{M-Theory Brane as Giant Graviton}
The $AdS_m\times S^n$ metric we adopt is \cite{GG2}:
\begin{eqnarray}
ds^2&=&ds_{AdS_m}^2+ds_{S^n}^2\label{ds2}\\
ds_{AdS_m}^2&=&-\left(1+\frac{r^2}{R^2_{AdS}}\right)dt^2+\frac{dr^2}
{1+\frac{r^2}{R^2_{AdS}}}+r^2d\Omega^2_{m-2}\label{dAdS}\\
ds_{S^n}^2&=&R^2_S(d\theta^2+\cos^2\theta d\phi^2+\sin^2\theta
d\tilde{\Omega}_{n-2}^2)\label{dS}
\end{eqnarray}
The coordinates on two spheres are separately denoted by
$\varphi_1$,...,$\varphi_{m-2}$ and
$\tilde{\varphi}_1$,...,$\tilde{\varphi}_{n-2}$. In the M-theory,
two cases are separately $AdS_4\times S^7$ and $AdS_7\times S^4$,
with the curvature radii separately satisfying $2R_{AdS}=R_S$ and
$R_{AdS}=2R_S$.\\
\indent Generally speaking, a giant graviton is a p-brane ($p\geq1$)
that have the same quantum number with the ordinary graviton but
extended in several spatial (either internal or outer) dimensions.
For simplicity we will mainly show the internal case, then a $p=n-2$
brane can wrap a $S^{n-2}$ in the internal $S^n$ space, while in the
uncompactified spacetime we see a worldline \cite{GG1, GG2}. Using
the diffeomorphism, the brane can be chosen to follow the static
gauge:
\begin{equation}
\xi_0=\tau=t\qquad\xi_1=\tilde{\varphi}_1\quad...\quad\xi_{n-2}=\tilde{\varphi}_{n-2}
\end{equation}
\begin{equation}
r=0\qquad\theta=\mbox{Constant}\qquad\phi=\phi(t)
\end{equation}
For outer noncompact space case we simply replace the condition by
$\theta=0,r=\mbox{Constant}$ and the spatial $\xi$s are the
$\varphi$s in the $AdS$ space. Similar giant graviton solution can
be constructed \cite{GG2}.\\
\indent For definiteness we consider $AdS_7\times S^4$ case, which
relates to the calculation we will perform in section 3. An M2-brane
wrap a $S^2$ in the internal $S^4$. The induced metric gives the
Nambu-Goto like action
\begin{eqnarray}
S_{NG}&=&-T_{M_2}\int_V d^3\xi\sqrt{-\det
\partial_iX^\mu\partial_jX^\nu
G_{\mu\nu}}\nonumber\\
&=&-4\pi T_{M_2}\int d\tau
R^2\sin^2\theta\sqrt{1-R^2\cos^2\theta\dot{\phi}^2}
\end{eqnarray}
where the $V$ denotes the worldvolume of the M2-brane.\\
\indent The $N$ coincident M5-branes which generate the AdS geometry
have background flux on the $S^4$. In the adopted coordinate system
the potential is given by
\begin{equation}
G^{(3)}_{\phi12}=R^3\sin^3\theta\sqrt{\det g_{\tilde{\Omega}_2}}
\end{equation}
where $g_{\tilde{\Omega}_2}$ is the metric on the internal $S^2$.
The pullback of the gauge potential gives the action
\begin{equation}
S_{gauge}=T_{M_2}\int_VG^{(3)}=4\pi T_{M_2}\int d\tau
R^3\sin^3\theta\dot{\phi}
\end{equation}
\indent So the overall lagrangian is given as
\begin{equation}
\mathcal{L}=4\pi T_{M_2}
R^2\sin^2\theta(-\sqrt{1-R^2\cos^2\theta\dot{\phi}^2}+R\sin\theta\dot{\phi})
\end{equation}
\indent From the action we can obtain some classical steady
configuration for giant graviton. The steps are the standard
procedure of classical mechanics, namely the Hamilton canonical
transformation. The Lagrangian doesn't contain $\phi$, which means
the canonical momentum $p_\phi$ is conserved. So the minimal energy
of fixed $p_\phi$ is determined only by the value of $\theta$
\begin{equation}
H=\frac{1}{R}\sqrt{p^2_\phi+\tan^2\theta(p_\phi-4\pi
T_{M_2}R^3\sin\theta)^2}
\end{equation}
The minimal energy is given by
\begin{equation}
\frac{\partial
H}{\partial\theta}\propto\frac{\sin\theta}{\cos^3\theta}(p_\phi-4\pi
T_{M_2}R^3\sin\theta)[p_\phi-4\pi
T_{M_2}R^3(2\sin\theta-\sin^3\theta)]
\end{equation}
One obvious solution is $\sin\theta=\frac{\displaystyle
p_\phi}{\displaystyle 4\pi T_{M_2}R^3}$. This means a nonvanishing
volume that the M2-brane takes in the $S^4$ is a steady
configuration. Given that all the quantum number of the membrane is
the same with a graviton, it is called a giant graviton, which also
indicates that the this configuration is a version of graviton that
has spatial extension.\\
\indent There is a further comment that we want to make. The ansatz
we adopted contains only an angle $\phi$ in the internal space $S^n$
as a function of the brane worldvolume variables, and the solution
we explicitly showed implies that the angular speed $\dot{\phi}$ is
a constant. The giant graviton rotates in a plane around some fixed
point. This is a strong evidence that while considering the giant
graviton as electrons and the rotation is due to the transverse
magnetic field, it is the Hall effect that the giant graviton
experiences \cite{GG1}. We will soon see that this effect extend to
the LLM case.

\subsection{The M-Theory Sector of LLM Geometry}
The original M-theory sector of AdS/CFT correspondence includes both
the $AdS_4\times S^7$ case and the $AdS_7\times S^4$ case, which are
separately the throat geometry of large number of M2-branes and
M5-branes. As we have seen, the giant graviton includes the same
brane spectrum for both cases. Namely, both the M2-brane and the
M5-brane can be implemented as giant graviton in either the
$AdS_4\times S^7$ case or the $AdS_7\times S^4$ case, whereas they
differ in whether the giant graviton is in the internal spheric
space or the outer $AdS$ space. So the LLM geometry describing them
can be uniformly obtained by requiring the $SO(3)\times SO(6)$
symmetry. Taking the $\frac{1}{2}$ BPS condition into consideration,
the geometry is given in \cite{LLM}
\begin{eqnarray}
ds^2&=&-4e^{2\lambda}(1+y^2e^{-6\lambda})(dt+V_idx_i)^2+\frac{e^{-4
\lambda}}{1+y^2e^{-6\lambda}}[dy^2+e^D(dx_1^2+dx_2^2)]\nonumber\\
& &+4e^{2\lambda} d\Omega_5^2+y^2e^{-4\lambda}d\tilde{\Omega}_2^2 \label{metricLLL}\\
G^{(4)}&=&F^{(2)}\wedge d^2\tilde{\Omega}_2\label{4formfield}\\
e^{-6\lambda}&=&\frac{\partial_yD}{y(1-y\partial_yD)}\\
V_i&=&\frac{1}{2}\epsilon_{ij}\partial_jD\qquad\mbox{or}\qquad
dV=d(V_idx_i)=\frac{1}{2}\ast_3[d(\partial_yD)+(\partial_yD)^2dy]\\
F^{(2)}&=&dB_t\wedge(dt+V)+B_tdV+d\hat{B}_t=d[B_t(dt+V)+\hat{B}_t]\\
B_t&=&-4y^3e^{-6\lambda}\\
d\hat{B}_t&=&2\tilde{\ast}_3[y^2(\partial_y\frac{1}{y}\partial_y
e^D)dy+ydx^i\partial_i\partial_yD]
\end{eqnarray}
where $i,j = 1,2$, $\ast_3$ is the Hodge star of the 3d metric
$dy^2+e^Ddx^2_i$ and $\tilde{\ast}_3$ is the 3d flat space Hodge
star. The function D which determines the solution obeys the 3d Toda
equation
\begin{equation}
(\partial^2_1+\partial^2_2)D+\partial^2_ye^D=0
\end{equation}
\indent Note that the coordinate $y$ is related to the radii of the
two spheres $S^2$, $S^5$ by $y=\frac{R_2R_5^2}{4}$ where
$R_2=ye^{-2\lambda}$ and $R_5=2e^\lambda$. So if the coordinate $y$
goes to zero, one of the two radii must go to zero and the function
$D$ is constrained in a subtle way in order to eliminate the
singularity. The condition is given in \cite{LLM}
\begin{equation}\label{nonsigular}
y\rightarrow0\Leftrightarrow\left\{
\begin{array}{*{2}{c@{\qquad}c}}
\partial_yD=0\quad D=\mbox{finite} & S^2\enspace\mbox{shrink}\\
D\sim\log y & S^5\enspace\mbox{shrink}
\end{array}\right.
\end{equation}
\indent Although generically, the nonlinear Toda equation can not be
analytically solved, two exact solutions are shown in \cite{LLM}.
They are based on the above nonsingular analysis and separately
correspond to the original AdS/CFT geometry. For the $AdS_4\times
S^7$ case
\begin{equation}
e^D=4L^{-6}\sqrt{1+\frac{r^2}{4}}\sin^2\theta\qquad
x=\left(1+\frac{r^2}{4}\right)^{\frac{1}{4}}\cos\theta \qquad
2y=L^{-3}r\sin^2\theta
\end{equation}
For the $AdS_7\times S^4$ case
\begin{equation}
e^D=\frac{r^2L^{-6}}{r^2+4}\qquad
x=\left(1+\frac{r^2}{4}\right)\cos\theta\qquad
4y=L^{-3}r^2\sin\theta
\end{equation}
where $r$ is a radial coordinate on $AdS$ and $\theta$ is an angle
on $S$, they function like the ones in (\ref{dAdS}) (\ref{dS}) but
are not exactly the same things. We transform the Cartesian
coordinates $x_1,x_2$ into polar coordinates and denote them by $x$,
$\tilde{\phi}$
\begin{equation}\label{transform}
ds^2_2=dx_1^2+dx_2^2=dx^2+x^2d\tilde{\phi}^2
\end{equation}
Only the $x$ appears in the above two formulae, which implies the
rotation symmetry. $L^{-1}$ is the curvature radius of $AdS_4(S^4)$
in the former(latter) case. The variables $r,\theta$ of the function
$D$ can be transformed into $x,y$ by the relation shown above. After
a few calculation, they give back precisely to the $AdS\times S$
geometry as shown in (\ref{ds2}), (\ref{dAdS}), (\ref{dS}). Some of
the steps go as follows
$$AdS_4\times S^7\qquad e^\lambda=\frac{\sin\theta}{L}\qquad
V_{\tilde{\phi}}=-\frac{4\cos^2\theta}{r^2+4\sin^2\theta}\qquad
\tilde{\phi}=\phi-t\nonumber$$
\begin{equation}
ds^2=\frac{1}{L^2}\left[-(r^2+4)dt^2+\frac{dr^2}{r^2+4}+\frac{r^2}{4}
d\tilde{\Omega}^2_2\right]+\frac{4}{L^2}[d\theta^2+\cos^2\theta
d\phi^2+\sin^2\theta d\Omega^2_5]
\end{equation}
$$AdS_7\times S^4\qquad e^\lambda=\frac{r}{2L}\qquad
V_{\tilde{\phi}}=\frac{2\cos^2\theta}{r^2+4\sin^2\theta}\qquad
\tilde{\phi}=2t-\phi\nonumber$$
\begin{equation}
ds^2=\frac{4}{L^2}\left[-(r^2+4)dt^2+\frac{dr^2}{r^2+4}+\frac{r^2}{4}
d\Omega^2_5\right]+\frac{1}{L^2}[d\theta^2+\cos^2\theta
d\phi^2+\sin^2\theta d\tilde{\Omega}^2_2]
\end{equation}
\indent The $AdS$ part of the geometry is not in the standard form
given in (\ref{dAdS}). However, perform a diffeomorphism rescaling
of $r\to2rL,\enspace t\to\frac{1}{2}tL$ for the former case and
$r\to rL,\enspace t\to\frac{1}{4}tL$ for the latter case, they
readily become the standard $AdS\times S$ geometry shown in
(\ref{ds2}). The relation between the curvature radii of the $AdS$
and $S$ parts is correctly reproduced.

\section{From Multi Giant Graviton to QHE Action}

\subsection{The Macroscopic Description}
Next we will consider the case of a number of M-theory branes in the
LLM geometry. The most direct way is, as used by \cite{GG1,GG2}, the
induce metric and the pullback of the gauge potential. As a
fundamental object, the induce metric is not explained as a kind of
Born-Infeld action that governs the low energy dynamics of D-branes,
but their appearance looks exactly the same. It is a direct further
generalization of the Nambu-Goto action of the least worldsheet area
\begin{eqnarray}
\mbox{M2-brane} &\qquad& S_{NG}=-T_{M_2}\int_Vd^3\xi\sqrt{-\det P[G_{ab}]}\\
\mbox{M5-brane} &\qquad& S_{NG}=-T_{M_5}\int_Vd^6\xi\sqrt{-\det
P[G_{ab}]}
\end{eqnarray}
where $P[G_{ab}]=\partial_aX^\mu\partial_bX^\nu G_{\mu\nu}$ is the
pullback of the metric on the worldvolume. $a,b$ label the
longitudinal directions along the brane worldvolume, and the left
transverse ones are denoted by $i,j$. Going to the static gauge we
have
$P[G_{ab}]=G_{ab}+G_{i\{a}\partial_{b\}}X^i+\partial_aX^i\partial_bX^jG_{ij}$.
\\
\indent Since the M2-brane electrically couples to the M-theory 3
form potential and the M5-brane magnetically, the pullback of the
gauge potential should also be included in the action. This gives
the Chern-Simons like part
\begin{eqnarray}
\mbox{M2-brane} &\qquad& S_{CS}=T_{M_2}\int_VC^{(3)}\\
\mbox{M5-brane} &\qquad& S_{CS}=T_{M_5}\int_VC^{(6)}
\end{eqnarray}
where $C^{(3)}$ is the gauge potential coupled to M2-brane and
$C^{(6)}$ coupled to M5-brane. They are related by
$\ast(dC^{(6)})=dC^{(3)}=G^{(4)}$.\\
\indent We will use the LLM frame rather than the $AdS\times S$
geometry to show the dynamics of the giant gravitons. The LLM
geometry, taking the $AdS\times S$ as a special case, will recover
the whole $\frac{1}{2}$ BPS sector which preserve certain bosonic
symmetries. For simplicity we will only consider the M2-brane case,
where we can directly use the 4-form field strength given in
\cite{LLM} (\ref{4formfield}) to determine the gauge potential. Note
that physically we ask the background LLM geometry to be
nonsingular, so what we are using is only taking the $S^5$ shrinking
condition (corresponding to the M2-brane case), i.e. factor 1 will
be ignored when accompanied by $y^2e^{-6\lambda}$, because this term
is in fact $\frac{1}{4}(\frac{R_2}{R_5})^2$. Since we are only
interested in the behavior of $y=0$ plane, the $e^D$ will be
replaced by its asymptotic value $y$. Then the Nambu-Goto part is
\begin{equation}
S_{NG}=-4\pi T_{M_2}\int dt[2y^3e^{-6\lambda}(1+V_i\dot{x}_i)-
\frac{1}{4}(\dot{x}_1^2+\dot{x}_2^2)]
\end{equation}
\indent The pullback of the 3 form part is
\begin{equation}
S_{CS}=2\pi T_{M_2}\int dt[4y^3e^{-6\lambda}(1+V_i\dot{x}_i)+
x_1\dot{x}_2-x_2\dot{x}_1]
\end{equation}
\indent So the combined Lagrangian is
\begin{equation}
\mathcal{L}=\frac{1}{2}(\dot{x}_1^2+\dot{x}_2^2)
+(x_1\dot{x}_2-x_2\dot{x}_1)
\end{equation}
\indent If we want to restore the dependence of the 11d M-theory
Planck length $\ell_P$, dimensional analysis gives the Lagrangian
\begin{equation}\label{oriLag}
\mathcal{L}=\frac{1}{2\ell_P}(\dot{x}_1^2+\dot{x}_2^2)
+\frac{1}{\ell_P^2}(x_1\dot{x}_2-x_2\dot{x}_1)
\end{equation}
It is exactly the one that describes the 2d charged particles which
is exposed to vertical magnetic field and subject to the Lorentz
force, namely, the one which governs the Quantum Hall Effect. We see
that the $x_1,x_2$ coordinates in the original LLM geometry in our
formulation is exactly 2d plane that the Quantum Hall Effect lives.\\
\indent For convenience of the treatment in the next subsection, we
will explore some aspect of the QHE side. Where our interest mainly
focus in, the Lowest Landau Level (LLL) captures the long distance
(or low energy) behavior of a QHE system. In the above case it is
equivalent to $\ell_P\rightarrow0$ limit, in other words, to omit
the kinetic part of the Lagrangian. Then only left with
\begin{equation}\label{LLLLag}
\mathcal{L}=x_1\dot{x}_2-x_2\dot{x}_1=x^2\dot{\tilde{\phi}}
\end{equation}
in the RHS we use the coordinate transformation (\ref{transform}).
Then we can see in the LLL limit it coincides with the original
ansatz of giant graviton. Only the derivative of the coordinate
$\tilde{\phi}$ appears in the Lagrangian and not the coordinate
itself, which means the symmetry along the $\tilde{\phi}$ direction.
Effectively
the $\tilde{\phi}$ is an isometric direction in this limit.\\
\indent One should not be surprised about the result. Going back to
the original giant graviton solution in $AdS\times S$, we can see
the rotation movement in $\phi$ direction. The above Lagrangian is
just an extension of this circumnutation. Based on this
consideration, we would argue that in the M5-brane case the
corresponding Lagrangian will be exactly the same, according to the
fact that it also contains the rotating giant graviton as a special
case.

\subsection{Towards the DLCQ Matrix Model in Weakly Curved Background}
In this subsection we will focus on the M2-brane case where the
microscopic description is somehow formulated. The above description
is purely macroscopic, which ignore the internal degree of freedom
of the M2-brane. The most crucial point that the transverse
direction $X_1,X_2$ should be matrices, is also
completely obscured.\\
\indent As we have noted in the introduction, it is believed that
the M2-brane takes the BFSS matrix model as a microscopic
description. In flat 11d spacetime and large $N$ limit the
description is exact \cite{BFSS}, while later Susskind \cite{DLCQ1}
have extended it. M-theory compactified on a light like direction
with $N$ units of KK momenta (discrete light cone quantization or
DLCQ) is described by the dynamics of $N$ D0-branes, i.e. 0+1
dimensional $U(N)$ Super Yang-Mills. In \cite{curvedM1} the
interaction between two matrix entities are computed at the
linearized order. With the addition of linearized supergravity
action, such interaction between two matrix model entities is
translated into the determination of the linearized supergravity
currents, and it is done in \cite{curvedM1}. Then viewed as a matrix
entity, an 11d light like graviton in some general background can be
treated in such formulism of the matrix model literature. Namely
given the 11d weakly curved spacetime and the flux, the additional
action of a grivton beyond the flat space matrix model action is
determined by the linearized supercurrent formula. In
\cite{curvedM2,curvedM3} the theory is directly formulated in
general weakly curved background, and it is shown that additional
linearized supergravity action is equivalent to the low energy
Born-Infeld and Chern-Simons action of D0-branes. Finally in
\cite{curvedM3} it is focused in the M-theory AdS/CFT geometry, a
microscopic description of M2-brane giant graviton is given. Giant
graviton in $AdS\times S$ geometry is described by a combination of
D0-branes as a fuzzy 2-sphere. This is related to the dielectic
effect studied by Myers \cite{Myers} which implies a lower
dimensional D(p-2n)-brane can couple to a higher dimensional RR
charge which originally should be taken by Dp-brane. We should check
carefully whether the techniques can be implemented in our
treatment.\\
\indent First of all, we must note that the DLCQ procedure requires
an isometric compact direction in which the 11d gravitation wave
propagates, then we can change our coordinate into light cone system
and discretize the Kuluza-Klein mode spectrum. On the D0-brane side
this direction is eliminated, in order to guarantee that the
D0-brane live in nine transverse direction. In the original
$AdS\times S$ case the giant graviton ansatz automatically satisfies
this requirement, it is the isometric direction $\phi$. At first
glance, in the LLM geometry (\ref{metricLLL}) we do not have such a
direction that is obviously the propagation direction of the KK
modes. However, it is not difficult to see that in the LLL limit of
the effective action (\ref{LLLLag}) the $\tilde{\phi}$ direction is
such a direction. That's because in the LLL limit and the coordinate
system after the transform (\ref{transform}), the system is
equivalently described by the ansatz
$X=\mbox{Constant},\tilde{\phi}=\tilde{\phi}(t)$, the dynamics are
the same. Physically in that limit all the movement of 11d
gravitation wave in other directions such as radial $x$ are all
small fluctuation that can be ignored. So the DLCQ still can be
carried out.\\
\indent On the D0-brane side the effective metric must be
introduced, which eventually eliminate one isometric direction so
that the D0-brane feels ten spacetime dimension. Meanwhile the gauge
potential in the M-theory case should be inner producted by some
transverse direction to give the dielectric effect for them to
couple to D0-branes. It is checked several times
\cite{curvedM2,curvedM3} that the action in the weakly curved
background is recovered by the nonabelian dielectric D0-brane
Born-Infeld action and Chern-Simons action, which is considered as
nontrivial test of the D0-brane matrix model description for 11d
M-Theory or its type IIA compactification. The checking is valid for
arbitrary background geometry and flux, so all the following
microscopic treatment and the matrix substitution are guaranteed.
For our purpose we will not repeat the checking procedure but use
the following action directly
\begin{eqnarray}
S_{BI}&=-&T_0\int
d\tau\mbox{Str}\left\{k^{-1}\sqrt{-P[E_{00}+E_{0i}(Q^{-1}-\delta)
^{ij}E_{j0}]det(Q^i_j)}\right\}\\
S_{CS}&=&T_0\int d\tau\mbox{Str}\left\{iP[(i_Yi_Y)C^{(3)}]\right\}
\end{eqnarray}
where $i=1,...,9$ label the transverse direction and in all
directions
$E_{\mu\nu}=\mathcal{G}_{\mu\nu}+k^{-1}(i_kC^{(3)})_{\mu\nu}$ where
$\mathcal{G}_{\mu\nu}=G_{\mu\nu}-k^2k_\mu k_\nu$ is the effective
metric. $k_\mu$ is the Killing vector of the isotropic spacelike
direction in which the 11d gravitation wave propagate and
$k=|k_\mu|$. In our case some coordinate transformation is needed
when we fix this direction, namely we can perform (\ref{transform})
and the $\tilde{\phi}$ is such a direction. The definition of
effective metric $\mathcal{G}_{\mu\nu}$ naturally subtracts the
contribution from the $k$ direction so that the D0-brane feels nine
transverse direction. We also take
$Q^i_j=\delta^i_j+ik[Y^i,Y^k]E_{kj}$. Finally $i_Y$ is the inner
product of gauge potential with the transverse Killing vector and
$(i_Yi_YC)_{\mu_1\mu_2...\mu_p}=Y^iY^jC_{ij\mu_1\mu_2...\mu_p}$. $Y$
labels arbitrary transverse direction and generally becomes matrix
valued.\\
\indent Corresponding to the giant graviton in \cite{curvedM3}, we
also need some fuzzy 2-sphere ansatz
\begin{equation}
Y^i=\frac{ye^{-2\lambda}}{\sqrt{\tilde{N}^2-1}}J^i
\end{equation}
where $i=1,2,3$ and $J^i$ form an $\tilde{N}\times \tilde{N}$
representation of $SU(2)$. Then
$(Y^1)^2+(Y^2)^2+(Y^3)^2=y^2e^{-4\lambda}\unity=R^2\unity$ implies
the fuzzy 2-sphere has the fixed radius. All the other directions
remain Abelian, i.e. proportional to $\unity$, the commutator
between them vanishes. This is because we only have a single
M2-brane as a whole, the transverse direction should not be matrix
valued. Note that corresponding to the LLL limit, we have eliminated
the propagation direction of the graviton wave, i.e. the only
coordinate which is function of time, so merely the steady
configuration can be studied in the above formulism and also the
Hamiltonian. We can not get explicitly the dynamical low energy
effective action. It can be parallelly shown that up to order
$\mathcal{O}(\frac{1}{N})$ the above microscopic description should
be coincident with the macroscopic way, i.e. treatment carried out
in the above subsection.\\
\indent However, this is only a single giant graviton which is
combined by a number of D0-branes. In the LLM geometry we are
interested in a number of $N$ giant gravitons, so we should revise
our ansatz into the block diagonal form
\begin{equation}
Y^i=\mbox{diag}\left(\frac{R_1}{\sqrt{\tilde{N}^2_1-1}}J^i_1,
\frac{R_2}{\sqrt{\tilde{N}^2_2-1}}J^i_2, \,.\,.\,.\,
,\frac{R_N}{\sqrt{\tilde{N}^2_N-1}}J^i_N\right)
\end{equation}
The matrices have $\sum^N_{n=1}\tilde{N}_n$ columns and rows. Our
aim is to treat each block along the diagonal as a size
$\tilde{N}_n\times\tilde{N}_n$ unit of the above giant graviton,
where $J^i_n$ is the $\tilde{N}_n\times \tilde{N}_n$ $SU(2)$
representation. In all we will have $\sum^N_{n=1}\tilde{N}_n$
D0-branes to combine into $N$ M2-brane giant gravitons.
Correspondingly we relax our requirement and do not ask the other
directions to be identity matrices, especially, in the $x$
direction of the (\ref{transform}).\\
\indent Now we are not interested in the microscopic side of each
gravitons, so we can ignore the $\tilde{N}_n\times \tilde{N}_n$
identity matrices contents and simply replace each of them for
number 1. Correspondingly in other effective transverse directions
we should also do this simplification, then each transverse
coordinate shrinks to $N\times N$ matrix to describe the $N$
M2-brane giant gravitons. This is equivalent to treat each giant
graviton as a particle, and ignore all the inner structure and
dynamics of the each of them. If the radial direction $X$ is still
diagonal, it is merely some superposition of noninteracting giant
gravitons, just the trivial ground state configuration. In general
it can be arbitrary matrix when the full dynamics is taken into
consideration.\\
\indent Now turn to the low energy effective action. Using the
correspondence between the microscopic and macroscopic description,
the Lagrangian corresponding to (\ref{oriLag}) is very simply
revised. The transverse coordinates $x_1,x_2$ should be replaced by
matrices, and correspondingly a trace should be added. We see from
the LLL limit (\ref{oldLLL}) that it is in fact only one effective
direction, namely radial $X$ in (\ref{transform}), so there should
be no commutator between the two $X_i$
\begin{equation}\label{orimatrix}
\mathcal{L}=\frac{1}{2\ell_P}\mbox{Tr}(\dot{X}_1^2+\dot{X}_2^2)
+\frac{1}{\ell_P^2}\mbox{Tr}(X_1\dot{X}_2-X_2\dot{X}_1)
\end{equation}
This is our starting point of the next formulation of the QHE
side.\\
\indent Finally let's make some comment on the M5-brane case.
Generally the microscopic description of the M5-brane is unclear, so
we can't perform such kind of microscopic analysis. But we can see
that in the M2-brane case, the only visible change from the
macroscopic description is the substitution of the coordinate $x_i$
by their matrix counterpart $X_i$, which is an indication of the
essential nonabelian property of a stack of M2-branes. At least
according to some dimensional reduction result in string theory
(say, the M5-brane double-dimensional reduction is the IIA string
D4-brane \cite{M5-D4}), the nonabelian property of transverse
direction should be inherited by the M5-brane from the D-brane side.
So corresponding to the macroscopic description, in the M5-brane
case we expect that the Lagrangian should also be revised as
(\ref{orimatrix}). Then the following formulation is also valid for
the M5-brane case.

\section{The Noncommutative Chern-Simons Description}
Recall that we want to perform a deep exploration on the Quantum
Hall Effect side of the correspondence, in order to shed light on
the uncovered aspect of the corresponding geometry. A well known
approach is the Noncommutative Chern-Simons theory description
\cite{QHNCCS1, QHNCCS2}, which will be our main guidance.\\
\indent First of all, we note that the matrix nature of the
coordinates implies that the system is fermionic. Since the
Lagrangian is invariant under $U(N)$ gauge symmetry, one of the
$X$s, say $X_1$, can be diagonalized by a gauge transformation. In
this eigenvalue basis, with notation $(X_1)_{mn}=\delta_{mn}x_{1n},
(X_2)_{mn}=y_{mn}, y_{nn}=x_{2n}$, a typical classical Lagrangian of
the matrix model reads
\begin{equation}\label{eigenvalue}
\mathcal{L}=\frac{1}{2\ell_P}\sum_{i,n}\dot{x}_{in}^2+\frac{1}{2\ell_P}\sum_{m\neq
n}\dot{\bar{y}}_{mn}\dot{y}_{mn}+\frac{1}
{\ell_P^2}\sum_n(x_{1n}\dot{x}_{2n}-x_{2n}\dot{x}_{1n})
\end{equation}
where $i=1,2$. The Hamiltonian description can be established in the
eigenvalue basis. However, quantum mechanically there is change of
measure in path integral from the matrix-element basis to the
eigenvalue basis, the Van der Monde determinant of the $X_i$
eigenvalue $\Delta(x_1)=\prod_{n<m}(x_{1n}-x_{1m})$ is inserted
\cite{Planar}. So the Hamiltonian in the quantum theory is given by
\begin{equation}
H=\frac{1}{\Delta(x_1)}\tilde{H}\Delta(x_1)
\end{equation}
where the $\tilde{H}$ on the RHS is the direct Legendre
transformation of the Lagrangian (\ref{eigenvalue}). Obviously the
eigenfunction $\Psi$ of the quantum Hamiltonian is related to the
original eigenfunction by $\Psi(x)=\Delta(x_1)^{-1}\tilde{\Psi}(x)$.
The minus sign which is produced by exchanging the two elements in
the Van der Monde determinant just indicates the fermionic nature of
the system. So not only the dynamics of a single giant graviton but
the statistics of the interaction system enable us to identify the
whole probe brane system as a realization of the Quantum Hall
Effect.\\
\indent From now on we will change the coefficient by
$\frac{1}{\ell_P}\rightarrow m$ as the effective giant graviton mass
and $\frac{1}{\ell^2_P}\rightarrow B$ as the magnetic field, in
order to borrow some concepts from the study of QHE. Note that we
have set the `electric charge' $e$ to unit. We are going to the
well-established Noncommutative Chern-Simons Description, which
means that we only focus on the long distance behavior. As mentioned
above, a \emph{prior} of doing so is to ignore the kinetic part of
the Lagrangian. Because all the phenomena we are interested in lie
in the LLL, in which the kinetic energy of the electron system is
degenerated into the lowest level \cite{QHE, LLL}, the kinetic part
can be dropped. On the M-theory side, this is equivalent to taking
the $\ell_P\rightarrow0$ limit, i.e., ignoring all the high order
corrections, which consists with the low energy limit. At the
moment, we must point out that there is a subtlety for doing this,
and postpone
treating it in a little later.\\
\indent There is something important that we can not ignore, namely
the fermionic statistics. Viewing the 2d electron gas as a
dissipationless fluid, the fermionic statistics is shown by the
property that the droplet can not be compressed. This is guaranteed
in the treatment of \cite{QHNCCS1} by introducing a potential $U$,
which can be viewed as the short-distance statistics effect and have
an equilibrium configuration when the droplet is not compressed. In
order to achieve this we can introduce another set of `comoving'
coordinates $y_i$, of which the graduation in the real space is
everywhere (gauge) adjustable, according to the real space
separation of the nearby electrons. In other words, the electron is
statically distributed on this `comoving' coordinates with even
density. The real space position is a map from the comoving
coordinates, and it depends on the time. After normalizing the
comoving coordinates density as $\rho_0$, the real space electron
density is given by the Jacobi $\rho=\rho_0\frac{\partial
y}{\partial x}=\rho_0\frac{\partial(y_1,y_2)}{\partial(x_1,x_2)}$.
The incompressibility is reflected by requiring that the Jacobi is
unity when the potential reaches its minimum and the configuration
is equilibrium. This is also called the Gauss Law Constraint. The
action is given by
\begin{equation}\label{oldLLL}
\mathcal{L}=\int
dy^2\rho_0\left[\frac{B}{2}\epsilon_{ij}\dot{x}_ix_j+U\left(\rho_0
\frac{\partial(y_1,y_2)}{\partial(x_1,x_2)}\right)\right]
\end{equation}
where we temporarily ignore the matrix nature of the $X_i$s.\\
\indent The theory has the area preserving diffeomorphism (APD)
symmetry. If we consider some transformation of the comoving
coordinates $y'_i= y_i+f_i(y)$ and requiring the Jacobi remains the
same, it can be factorized as $\frac{\partial y'}{\partial
x}=\frac{\partial y}{\partial x}\frac{\partial y'}{\partial y}$ and
the solution is readily obtained
\begin{equation}
f_i=\epsilon_{ij}\frac{\partial\Lambda(y)}{\partial y_j}
\end{equation}
So the induced transformation of the $x_a$ will be
\begin{equation}
\delta x_a=\frac{\partial x_a}{\partial
y_i}f_i(y)=\epsilon_{ij}\frac{\partial x_a}{\partial
y_i}\frac{\partial\Lambda(y)}{\partial y_j}
\end{equation}
\indent Obviously there is a trivial solution to the equilibrium
configuration, i.e., we can choose the comoving coordinates such
that $x_i=y_i$. While we are not interested in this trivial
solution, we can perform a deformation to obtain new interesting
solutions, using the above APD symmetry technique. For instance
\[
x_i=y_i+\epsilon_{ij}\frac{1}{2\pi\rho_0}
\]
\indent The most direct way to achieve this APD is by introducing an
auxiliary field $A_0$, and requiring the equation of motion of $A_0$
gives the desired Gauss Law Constraint. The action is constructed in
\cite{QHNCCS1}
\begin{equation}
\mathcal{L}=\int
dy^2\left\{\frac{B\rho_0}{2}\epsilon_{ij}[(\dot{x}_i-\frac{1}{2\pi
\rho_0}\{x_i,A_0\})x_j+\frac{\epsilon_{ij}}{2\pi\rho_0}A_0]-\frac{1}{2}\rho_0m\omega^2x_i^2\right\}
\end{equation}
where classically the commutator $\{\enspace,\enspace\}$ is defined
as the Poisson Bracket with derivatives to $y_i$. The variation of
the auxiliary field $A_0$ gives the constraint
\begin{equation}
\{x_1, x_2\}=1
\end{equation}
This is exactly the inverse version of the equilibrium condition,
from which the APD symmetry can also be deduced. There is no
constraint on the value of $A_0$. For simplicity we can even choose
the gauge $A_0=0$, then the new Lagrangian goes back to
the old one (\ref{oldLLL}).\\
\indent We also add a term $-\frac{1}{2}\int
dy^2\rho_0m\omega^2x_i^2$ in the Lagrangian, where
$\omega=\frac{B}{m}$ is the cyclotron frequency. Recall that the
frequency also equals to the reciprocal of the 11d Planck length
$\ell_P^{-1}$ from our original QHE description (\ref{orimatrix}),
so we can see that the LLL condition $\omega \to \infty$ indeed
match the low energy limit of the M-theory side $\ell_P \to 0$ very
well. It is the subtlety we pointed out above. Let us see the reason
of doing this. What we are interested in is the `droplet' solution
that all the giant gravitons concentrate. The fermionic exclusion
principal as well as the fixed cyclotron frequency gives us a
picture that classically the droplet is a rigid body made up of
incompressible fermions that experience an overall circumnutation.
What is more, in fact we do not treat the rotation movement
explicitly in the following. In this sense going to the LLL can be
interpreted as the frame of reference transformation, i.e. we are
going to the rotating frame and the giant gravitons seem static to
combine into a droplet. From classical mechanics we know that when
transformed into a noninertial frame an inertia force must be
introduced, which is equivalent to the potential we added. From the
droplet constraint we notice that increasing the radius means more
energy, in this way the sign of the term can be determined.\\
\indent Right now we treat the real space coordinates as ordinary
number, and this is equivalent to treating the system as uniform
continuous fluid. But indeed they are some finite number of giant
gravitons, which have the discrete property. So it is high time that
we replace the simplified $x_i$ by their $N\times N$ matrices
counterparts $X_i$, meanwhile the Poisson bracket gives way to
matrix commutator, integral replaced by matrix trace and the
coefficient should be revised
\begin{equation}\label{enough}
\mathcal{L}=\frac{B}{2}\mbox{Tr}\{\epsilon_{ij}(\dot{X}_i-i[X_i,
A_0])X_j+2\theta A_0-\omega X_i^2\}
\end{equation}
where $\theta=\frac{1}{2\pi\rho_0}$. The continuous comoving
coordinate $y$s are replaced by the comoving lattices which are
mapped into the real space matrix element. Recall that the dynamics
we obtained in (\ref{orimatrix}) is already in terms of matrix, so
it is the matrix version that just corresponds to the original
action.\\
\indent The relationship between this theory and the standard
noncommutative Chern-Simons theory is elaborated in \cite{QHNCCS1,
NCCSNCV}. Ignoring the centrifugal potential, the standard
Noncommutative Chern-Simons form is obtained
\begin{equation}
\mathcal{L}_{NC}=\frac{1}{4\pi\nu}\epsilon_{\mu\nu\rho}\left(\hat{A}
_\mu\ast\partial_\nu\hat{A}_\rho+\frac{2i}{3}\hat{A}_\mu\ast\hat{A}_\nu\ast\hat{A}_\rho\right)
\end{equation}
But we will not explore further in such pure noncommutative
Chern-Simons theory direction. In fact, knowing (\ref{enough}) is
enough for our formulation.\\
\indent There are two different commutativity in the theory. One
originates quantum mechanically, the canonical conjugation of the
coordinate matrix is $\Pi_i=\frac{B}{2}\epsilon_{ij}X_j$ so the
canonical quantum condition gives
\begin{equation}\label{commutator}
[X_1,X_2]_{QM}=\frac{i}{B}
\end{equation}
The other is the APD symmetry constraint from the equation of motion
of the auxiliary field $A_0$
\begin{equation}
[X_1,X_2]_{APD}=i\theta
\end{equation}
In the following we will see that the ratio of the two commutative
parameters is related to the filling factor of the FQHE.

\subsection{Finite Noncommutative Chern-Simons and the Edge
Excitation}
However, the matrix version has some intrinsic
inconsistence, which must be revised. Recall that the matrix order
$N$ is the number of giant gravitons, which is large but finite. The
problem is that it is impossible to satisfy the APD symmetry in the
finite $N$ case. So the action should be revised by adding the edge
excitation \cite{QHNCCS2}. We must admit that in the M-theory side,
because our ignorance of the microscopic dynamics of the theory
itself, a satisfactory microscopic picture is still lacking.
\begin{equation}
\mathcal{L}=\frac{B}{2}\mbox{Tr}\{\epsilon_{ij}(\dot{X}_i-i[X_i,
A_0])X_j+2\theta A_0-\omega X_i^2\}+\Psi^\dag(i\dot{\Psi}-A_0\Psi)
\end{equation}
where the introduced field $\Psi$ lives only on the edge of the
droplet which is formed by finite $N$ discrete giant gravitons. The
term of $\Psi$ looks like the Dirac field the dynamics of which is
the first order equation, and correspondingly, it lives in the
fundamental representation of the $SU(N)$ Lie algebra. The $X_i$s
are in the adjoint representation. Namely, the symmetry
transformation is
\begin{equation}
X_i\rightarrow UX_iU^{-1}\qquad\Psi\rightarrow U\Psi
\end{equation}
\indent The extended Gauss Law Constraint is still the equation of
motion of the auxiliary field $A_0$. It is
\begin{equation}\label{Gauss}
G=iB[X_1,X_2]-\Psi\Psi^\dag+B\theta=0
\end{equation}
Taking the trace of the above equation gives
\begin{equation}
\Psi^\dag\Psi=BN\theta
\end{equation}
Note that in the finite $N$ case where the trace of commutator of
finite dimensional matrices is zero, if we do not introduce the
$\Psi$, doing the same steps above immediately gives us the
inconsistency. This explains the introduction of $\Psi$. The
equation of motion for $\Psi$ in the $A_0=0$ gauge is
$\dot{\Psi}=0$, so we can take it to be
\begin{equation}
\Psi=\sqrt{BN\theta}|\upsilon\rangle
\end{equation}
where $|\upsilon\rangle$ is a constant vector of unit length. Then
the traceless part of (\ref{Gauss}) reads
\begin{equation}\label{constraint1}
[X_1,X_2]=i\theta(1-N|\upsilon\rangle\langle\upsilon|)
\end{equation}
\indent The equation of motion of the $X_i$ field is
$\dot{X}_i=\omega\epsilon_{ij}X_j$. This is just a matrix oscillator
and solved by
\begin{equation}
X_1+iX_2=e^{i\omega t}A
\end{equation}
where $A$ is any $N\times N$ matrix satisfying
\begin{equation}\label{constraint2}
[A,A^\dag]=2\theta(1-N|\upsilon\rangle\langle\upsilon|)
\end{equation}
\indent To find the ground state of the system, we must minimize the
potential
\begin{equation}
V=\frac{B\omega}{2}\mbox{Tr}(X_1^2+X_2^2)
\end{equation}
with the constraint (\ref{constraint1}) or (\ref{constraint2}). This
can be done with
\begin{equation}
A=\sqrt{2\theta}\sum_{n=0}^{N-1}\sqrt{n}|n-1\rangle\langle
n|\qquad|\upsilon\rangle=|N-1\rangle
\end{equation}
This is essentially a quantum harmonic oscillator and hamiltonian
projected to the lowest $N$ energy eigenstates. It is easy to check
that the above satisfies (\ref{constraint2}). It represents a
circular quantum Hall droplet of radius $\sqrt{2N\theta}$. The
radius squared matrix coordinate $R^2$ is
\begin{eqnarray}
R^2&=&X_1^2+X_2^2=\frac{1}{2}(A^\dag A+AA^\dag)\nonumber\\
 &=&\sum^{N-2}_{n=0}\theta(2n+1)|n\rangle\langle
 n|+\theta(N-1)|N-1\rangle\langle N-1|
\end{eqnarray}
The highest eigenvalue of $R^2$ is $(2N-1)\theta$. So the giant
graviton density is $\rho=\frac{N}{\pi
R^2}=\frac{1}{2\pi\theta}=\rho_0$, gives back to the original
density. Classically they rotate around the origin with the
frequency $\omega=\frac{B}{m}=\ell_P^{-1}$.\\
\indent This is the ground state of the system. For excitation state
we are interested in the `quasiparticle' and `quasihole' states. For
a quasihole of charge $-q$ sets at the origin the solution is given
in \cite{QHNCCS2} as
\begin{equation}
A=\sqrt{2\theta}\left(\sqrt{q}|N-1\rangle\langle0|+\sum_{n=1}^{N-1}\sqrt{n+q}|n-1\rangle\langle
n|\right)\qquad q>0
\end{equation}
The required condition (\ref{constraint2}) can also be checked
explicitly. Meanwhile, the coefficient of the matrix
$|n\rangle\langle n|$ goes like $2\theta(n+q)$, so the lowest
nonzero mode is (approximately) $q$ and this is the picture of a
quasihole. And now the outer radius of the droplet is shifted to
$\sqrt{2(N+q)\theta}$. A number of $q$ giant gravitons in the center
of the droplet are excited to the outer of the droplet, and
explicitly the area of the droplet remains the same, which is just
the APD requirement and the reflection of the Fermion nature.\\
\indent Finally, the most general solution that can be viewed as
giant gravitons excitation is
\begin{equation}
A=\sqrt{2\theta}\sum_{i=1}^m\left(\sqrt{q_i}|n_i\rangle\langle
n_{i-1}|+\sum_{n=n_{i-1}+1}^{n_{i-1}}\sqrt{n+q_i}|n-1\rangle\langle
n|\right)
\end{equation}
where $|n_0\rangle=|0\rangle$ and $|n_m\rangle=|N-1\rangle$. One can
see that it describes $m$ groups of giant gravitons excitation.

\subsection{Quantization and the Fractional Quantum Hall Effect}
We now come to the question of quantization of the above matrix
model. After obtaining the matrix eigenstates, now we have no reason
to neglect the matrix effect and the canonical quantization
condition (\ref{commutator}) becomes
\begin{equation}
[(X_1)_{mn},(X_2)_{kl}]=\frac{i}{B}\delta_{ml}\delta_{nk}
\end{equation}
or in terms of $A=X_1+iX_2$
\begin{equation}
[A_{mn},A^\dag_{kl}]=\frac{1}{B}\delta_{mk}\delta_{nl}
\end{equation}
\indent Now we have the new edge excitation field $\Psi$ and its
quantization should also be included. We quantize it as boson
\begin{equation}
[\Psi_m,\Psi_n]=\delta_{mn}
\end{equation}
\indent So the system is a priori just $N(N+1)$ uncoupled
oscillators. What couples the oscillators and reduces the system to
effectively $2N$ phase space variables (the planar coordinates of
the giant gravitons) is the Gauss law constraint (\ref{Gauss}).
Following the treatment of \cite{QHNCCS2}, we show note that $X$,
$\Psi$ terms ($G_X$, $G_\Psi$) in (\ref{Gauss}) is exactly the
quantum generator of the $U(N)$ algebra. From group representation
theory we know that if $R^a_{\alpha\beta}$ is the matrix element of
the generators of $SU(N)$ in any representation, and
$a_\alpha,a_\beta$ a set of mutually commuting oscillators, then the
operator
\begin{equation}
G^a=a^\dag_\alpha R^a_{\alpha\beta}a_\beta
\end{equation}
satisfy the $SU(N)$ algebra. Recall that $X_i$ lives in the adjoint
representation and $\Psi$ in the fundamental, the equivalent form of
$G_X$, $G_\Psi$ in terms of oscillator basis is
\begin{equation}
G^a_X=-ia^\dag_bf^{abc}a_c\qquad G^a_\Psi=\Psi^\dag_mT^a_{mn}\Psi_n
\end{equation}
where the $f^{abc}$ is the structure constant of the $SU(N)$ Lie
algebra. So finally, in the form of acting on physical states, the
constraint that expressed separately in traceless and trace part are
\begin{eqnarray}
\mbox{traceless}&\qquad&(G^a_X+G^a_\Psi)|phys\rangle=0\\
\mbox{trace}&&(\Psi^\dag_n\Psi_n-BN\theta)|phys\rangle=0\label{trace}
\end{eqnarray}
\indent Focusing on the trace part, we notice that the former part,
the occupation number operator $\Psi^\dag_n\Psi_n$ takes integer
eigenvalues, so the $BN\theta$ should also take integer value. What
is more, the traceless equation shows that the physical state is in
a singlet representation of the $SU(N)$. Since physical states are
invariant under the sum of $G_X$ and $G_\Psi$, the representations
of $G_X$ and $G_\Psi$ must be conjugate to each other, so their
product contains the singlet. Therefore, the irreducible
representation of $G_\Psi$ must also have a number of boxes in Young
tableau which is a multiple of $N$. The oscillator realization
(\ref{trace}) contains all the symmetric irreducible representation
of $SU(N)$, whose Young tableau consists of a single row. The number
of boxes equals the total number operator of the oscillators
$\Psi_n^\dag\Psi_n$. So we conclude that $BN\theta$ must be an
integer multiple of $N$, that is
\begin{equation}
B\theta=\nu^{-1}=k\qquad k=\mbox{integer}
\end{equation}
This is nothing but the FQHE inverse filling factor quantization
condition. The reciprocal $\nu=k^{-1}$ is the more used filling
factor. Classically the integer can be any nonnegative integer, but
quantum mechanically there is a shift $k\rightarrow k+1$, which can
be viewed as the effect of zero point energy \cite{refer}. So the
inverse filling factor is strictly positive. Further from standard
argument in QHE we know that the odd $k$ corresponds to the
\emph{fermionic} statistics and the even the \emph{bosonic}. Given
the fermionic statistics, we get that the $k$ should be odd numbers.
So finally we obtain the fractional QHE behavior with the odd
inverse filling factor.

\subsection{The Effective Interaction of the Giant Graviton}
To determine the interaction in the generalized Quantum Hall Effect,
it is convenient to adopt the Hamiltonian approach. The Hamiltonian
of the system is
\begin{equation}
H=V=\frac{B\omega}{2}\mbox{Tr}X_i^2=\frac{B\omega}{4}\mbox{Tr}(AA^\dag+A^\dag
A)
\end{equation}
At the classical level, the Gauss law constraint can be solved
\cite{QHNCCS2} in the eigenvalue basis of $X_1$
\begin{equation}
(X_1)_{mn}=x_n\delta_{mn} \qquad
(X_2)_{mn}=y_n\delta_{mn}+\frac{i\theta}{x_m-x_n}(1-\delta_{mn})
\end{equation}
Substituting the solution into the classical Hamiltonian, one
obtains the Hamiltonian in terms of the variables $x_n$
\begin{equation}
H=\sum_{n=1}^N(\frac{\omega}{2B}p_n^2+\frac{B\omega}{2}x_n^2)+\sum_{n\neq
m}\frac{k(k-1)}{(x_m-x_n)^2}
\end{equation}
The latter term describes the interaction potential between
different giant gravitons. It is nothing but the integrable
one-dimensional Calogero model \cite{QMBM} for non-relativistic
particles on a line.\\
\indent We see that going back to the IQHE $k=1$ the interaction
term vanishes. So the interaction is a novel property of the FQHE.
We can also see the Hamiltonian imply that interaction between the
giant gravitons is \emph{repulsive}, for compressing the system
means greater potential. This is common in the condense matter
physics.\\
\indent The ground-state wave function of the Calogero model
\begin{equation}
\Psi_0(x_1,x_2,...)=\prod_{m<n}(x_m-x_n)^k\exp\{-\frac{B}{2}\sum_nx_n^2\}
\end{equation}
is nothing but the 1d representation (in the Landau gauge) of the
Laughlin wave function in the LLL with $\nu=k^{-1}$ on a disk
geometry. In this way the relationship between the $k$ and the
statistics can be clearly seen.

\section{Conclusion and Speculation}
We have explicitly shown that at the probe limit, the M-theory
sector of the AdS/CFT/QH correspondence in the LLM geometry can
adopt Fractional QHE. In fact, the low energy effective action of
the probe giant graviton in the LLM geometry is exactly the QHE,
this is enough for all the later derivation. Because the original
correspondence \cite{LLM} only relates the Integer QHE sector (for
example, see \cite{LLL}), finding the exact counterpart in the
supergravity side and the CFT side should be an interesting open
question.\\
\indent Although the dual geometry is an interesting open question,
we want to make some speculation on it. In the IIB case, the
nonsingular condition requires the function $z$ to be
$\pm\frac{1}{2}$. After a simple shift $\tilde{z}=z-\frac{1}{2}$ the
2d configuration is truly black-white, so the fractional particle
density form the fractional QHE side is obviously identified as
$\tilde{z}$. However, in the M-theory sector the nonsingular
condition (\ref{nonsigular}) is far from directly identified as a
fermionic density-like function that adopts two values, which makes
the fractional density explanation difficult. To attack the problem,
corresponding to the IIB case, we can make an intuitive
identification between the two nonsingular condition and the
occupied/unoccupied states of the giant gravitons. For example, we
can put the $S^5$ shrinking condition in the M2-brane case
corresponding to the unoccupied state as the background. Then we
speculate that the fractional charge density requirement from the
fractional QHE side acting on the function $D$ just corresponds to a
state different from and in some sense between the two $D$s that
satisfy the separated nonsingular condition. A direct linear
combination, say $D=\rho D_1+(1-\rho)D_2$ where $D_1$ and $D_2$
satisfy separately the $S^2$ and $S^5$ shrinking condition of
(\ref{nonsigular}), will not work. That's because the required Toda
equation is not a linear equation. Needless to say, in our
explanation the geometry of such kind of solution, if translated by
the standard LLM dictionary \cite{LLM}, is \emph{singular}. This is
just because the obvious deviation from the nonsingular condition.

\end{document}